# From Prohibition to Adoption: How Hong Kong Universities Are Navigating ChatGPT in Academic Workflows

Junjun Huang[1,2], Jifan Wu[1], Qing Wang[1], Kemeng Yuan[1], Jiefeng Li[1], Di Lu[1]

*1. Department of Management and Marketing, Faculty of Business, The Hong Kong Polytechnic University, Hung Hom, Kowloon, 999077 Hong Kong.*

*2. Department of Artificial Intelligence, Faculty of Computer Science and Information Technology, Universiti Malaya, 50603 Kuala Lumpur.*

## ABSTRACT

This paper aims at comparing the time when Hong Kong universities used to ban ChatGPT to the current periods where it has become integrated in the academic processes. Bolted by concerns of integrity and ethical issues in technologies, institutions have adapted by moving towards the center adopting AI literacy and responsibility policies. This study examines new paradigms which have been developed to help implement these positives while preventing negative effects on academia.

# 1. Introduction

Due to the emergence of ChatGPT and other generative AI tools recently in the market has caused a shift in the traditional academic environment hence varying reactions from universities across the globe. As for the universities In Hong Kong, the response towards ChatGPT was initially quite skeptical with outright bans being implemented since there were issues of academic dishonesty and ethical concerns. However, once its educational capabilities have been realized, institutions started transitioning to a rather regulated approach, primarily concerned with creating guidelines for the practical utilization of these technologies that would address concerns of ethicality, but also creativity.

While ChatGPT is banned initially by most of the eight major universities in Hong Kong, some universities adopt the new tech with more careful and regulated policies. At first, the University of Hong Kong and the Chinese University of Hong Kong prohibited the use, but now permit its use subject to specific guidelines on academic integrity. Hong Kong Baptist University is one of the other universities that have stricter bans to avoid academic dishonesty. Now most schools allow the use of GPT, providing there is proper supervision and transparency that the tool complements the learning and not undermines it.

When the eight universities in Hong Kong implemented their own policies on ChatGPT, they diverged, with some banning use because of academic integrity. Eventually, almost all institutions began to adopt a regulated approach to tackle innovation and ethical academic practices with training in AI literacy. This change follows a larger movement in society to realize that AI is a tool for academia to liberate itself from monotony and become more efficient and more creative. However, integrating an AI tool such as GPT requires frameworks that help maintain the role of AI as a complement to – and not a substitution or destroyer of – critical thinking, nor encourage academic dishonesty. The adoption represents efforts universities are making towards a responsible use of AI, while retaining high academic standards.

Theoretically, this transition supports Innovation Diffusion Theory (Rogers, 1962) that describes the way in which new technologies diffuse within societies. At first sight, ChatGPT is dismissed, but universities are now transitioning from rejecting ChatGPT on the grounds of potential misuse to adopting it systematically via AI literacy programs as a tool for academic enhancement. The practice indicates that due to the gradual adoption, the theory's stages can be used to illustrate — awareness, interest, evaluation, and acceptance — the educational institutions, like the others, is quite cautious, but optimistic, in its adoption of technologies. Hong Kong universities hope to strike a balance between promoting innovation and promoting ethical use.

This paper aims to analyze the development of the use of chat GPT in Hong Kong universities from banning to systematic use and the effects on the academic integrity and AI awareness and knowledge.

## 2. Initial Prohibition and Ethical Challenges

At first, the appearance of ChatGPT in Hong Kong's universities dealt with restrictive measures, which are also observed on an international level. Of the participants' concerns list, academic dishonesty such as plagiarism, fabrication of data, and the unauthorized application of AI in academic activities stood out (Chan, 2023; Lee & Chen, 2024). Such concerns were not for naught as that generated by AI was highly likely to compromise academic integrity and worth of achievements within academic community. Echoing Banipal, Asthana and Mazumder (2023), they pointed out the lack of strong ethical framework resulted into these issues forcing institutions to temporarily or permanently block ChatGPT as a preventative measure. As shown in Table 1, the initial focus of the prohibition was to address academic misconduct such as plagiarism and data fabrication, but over time, university policies shifted towards introducing AI literacy programs to address these issues.

**Table 1: Summary of Ethical Concerns and Policy Responses in Hong Kong Universities**

| Ethical Concerns | Initial Response | Current Policy Approach | Remarks |
|---|---|---|---|
| Plagiarism | Ban on AI use in assessments | Integration of AI literacy programs | Ongoing monitoring needed |
| Fabrication of Data | Restrictive policies | AI literacy with strict usage guidelines | Guidelines under review |
| Unauthorized AI Application | Full ban | AI ethics and responsibility training | Improvement in compliance |

However, the use of AI in education was attended ethical issues with regards to its sustainability was also an issue of blind spots in policy-making as observed by Driessens and Pischetola (2023). Academic institutions tried to find a balance between growth that comes with technology and maintaining the 'purity' of education, and mostly, the latter was sacrificed. It is these early prohibitive measures that also touched on a more general ethical dilemma of the possibility of the technology versus the maintaining of academic integrity.

## 3. Transitioning Towards Adoption: Policy Reforms and AI Literacy

However, given several restrictions in the course of this study, it was possible to establish more redeeming qualities for using ChatGPT and improving academic practices (Rasmussen & Karlsen, 2023). They have started to change their stance to the policies that they have adopted to be more restrictive and shift towards the management of AI in a sustainable manner by Hong Kong universities. As pointed out by Bhullar, Joshi, and Chugh (2024), the transformation is apparent pointing to the fact that later on, the institutions began formally incorporating the AI guidelines into the academic policies shifting the gear from an outright ban to AI literacy and ethical use.

To support this shift, the emergence of the systematic AI policy frameworks served as this transition's cornerstone. In Chan (2023) model, student and educator acquire the necessary skills to enable them tackle the ethical and practical issues of generative AI. This approach reflects the global shift since more institutions are adopting the need to teach students on the ethics of AI and how they can utilize it responsibly as part of the curriculum (Jiao et al., 2024; Ferdaus, et al., 2024). Through promoting a deep understanding of the nature and workings of AI and AI systems, universities in Hong Kong are prepared to enhance the awareness and the critical thought level of stakeholders in order to ethically address and manage AI technologies and their risks of the academic misconduct.

## 4. Integration and Future Implications

Substitution and integration have occurred incrementally and in limited ways, and we have seen the mixed benefits in the use of ChatGPT in academic contexts. According to Bhaskar and Gupta (2024), it should be important to look at AI as an augmentation rather than a replacement of the human decision in learning contexts. For example, ChatGPT has applied to management education where it has been adopted to present examples of realistic management decision-making situations that give students exposure to as many positions as possible, so that they emerge with sharper thinking skills (Mai, Da & Hanh, 2024; Soubhari, et al., 2023).

Nevertheless, the complete integration seems to be possible, but it has many ethical concerns. ChatGPT's role in higher education in the future will then rely on the establishment of good ethical standards that would meet the emerging ethical issues associated with the application of AI. In Murgatroyd (2024), the author points out that such politics require constant improvement and propose the use of artificial intelligence to achieve the educational agenda without compromising on integrity. While Universities grapple with the complexity of this environment ethical AI literacy and responsible implementation becomes the key to realizing the potential of generative AI in universities (Olney, 2024; Xu, 2024).

## 5. Conclusion

The transition of Hong Kong universities from banning ChatGPT to recognize it as an essential pedagogical assistant is a microcosm of the transition happening to the meta-narrative of higher learning. Such a transition calls for policy parity which seeks to educate the public on AI applications, how it can be utilized in academic institutions and ensure the application of higher standards of education as and when it is needed. As such, with the heightened emphasis on appropriate AI utilization in the years to come, the policies will affect the direction of academic processes in terms of potential development and adjustment to the newcomer's trends.

## 6. Team Reflections and Case Studies on AI Policy in Hong Kong Universities

In this section, we explore various viewpoints from team members who have firsthand experience and insights into how Hong Kong universities are addressing the challenges and opportunities of integrating AI tools, such as ChatGPT, into academic workflows. Each of the reflections provided highlights different facets of the AI adoption journey, contributing to a more comprehensive understanding of the evolving academic landscape.

- AI as an Augmentation Tool

Li Jiefeng argues that as AI gradually becomes an integral part of our daily lives, it is more important to be proponents of this new order rather than opponents of this inevitable change. In academic sphere AI such as ChatGPT is very useful for students by opening them up new opportunities in creativity and increasing access to academic materials for the teachers. AI integration may prompt thinking in the use of new approaches during the problem-solving process and creativity that is important in the current educational practice.

However, the openness and acceptance of AI in academic has been welcomed greatly, yet this has given rise to a number of problems regarding integrity in academics. A definite disadvantage of using such services is the possibility of their misuse, for example, in plagiarism cases or when involving AI in writing too many papers. Thus, to enable applying the benefits of AI technology in enhancing the learning process, it is necessary to find a balance between increasing the effectiveness of the learning process and diminishing the negative impacts that might occur due to the use of AI. It can be unbalanced if there is no effective AI literacy programs that will educate both the students and the faculty on how these tools can be exploited in the right manner. Thus, academic institutions will be able to promote AI as a tool enhancing human creativity and not a tool replacing it when they help people develop a better understanding of AI's strengths and weaknesses.

Moreover, educators are in a position where they need to adopt certain policies resulting in the use of AI responsibly. Universities should thus come up with clear guidelines and policies to protect academic integrity since the underlying technologies are very useful in the creation of new value. This proactive approach will discipline the academic community to handle the realities of AI integration effectively, thereby making the use of such technologies to be in harmony with the proper and noble objectives of learning, creating, and being ethical in practice.

- Balancing Innovation with Critical Thinking

Yuan Kemeng suggests that in order to cultivate talents with innovative thinking and independent skills, future universities must change their educational methods. As AI technology evolves rapidly with tools such as ChatGPT students achieve more efficient data retrieval and improve their research and language skills. However, this convenience poses a risk: if students depend too heavily on AI, they might undermine their progress in critical thinking and problem-solving abilities necessary for managing a complex world.

AI ought to be recognised as a strong asset that supports instead of substitutes human intelligence. Although AI can boost academic performance for students, they should also cultivate their own abilities and refrain from unthinkingly reproducing content from AI. Universities guide students to find this equilibrium. By embedding AI in the classroom and guiding students to use it in a responsible manner institutions can enable access to its advantages while developing self-sufficiency and creativity.

AI should be viewed as a powerful tool that augments rather than substitutes human cognition. Certainly, students use AI to augment their scholarly work, but they must also build capacity and adapt as they should not copiously just replicate what AI produces. It is the duty of universities to help students find this balance in their lives. Ensuring that students have the proper tools to responsibly use AI in the classroom will help them realize the positive aspects of AI while also keeping their skills more well-rounded and creative.

For the Open Review of AI, as AI will become a core part of our future society, universities have to be at the forefront in order to gear students on how they can use it wisely. It depends on not only training technically but also changing cultural mindsets towards independent thinking, critical analysis and innovation. Thus, the academic community can undermine AI's potential without any cost of losing important human capacities for the future.

- The Risk of Data Integrity

Wang Qing notes that, due to the expansive AI technologies like GPT, we are now exposed to more data simultaneously than at any previous time in history. Although this dramatic increase in the ability to rapidly access data has greatly improved research efficiency, it also presents new issues around the precision and thoroughness of data obtained. It is very fast in creating the content; however, outputs are not reliable with that due to GPT would have been trained on vast datasets which may include misguiding information rejected from many articles. This excess of unchecked, non-credible data presents the temptation to incorporate incorrect information into studies, which will lead to discovery invalidity.

GPT can have a lot of benefits for the humanities, that is speed up literature review, data synthesis or language recourse — nonetheless you should be warned: for academic work. Blindly using AI and believing in the power of this tool can jeopardize the veracity of your research. It needs to be more balanced and nuanced; users need to double-check what AI says through one or two of the correct sources, then they will find a dearth of facts.

In addition, universities and researchers should enforce strong standards that ensure the ethical use of AI in research. However, scholars should not be complacent with all outputs from AI models like GPT but rather they are supposed to practice their critical thinking and rational judgment for what is right or acceptable information they use. But when it comes to original scholarship, AI—especially in its present state—is just one weapon in the scholar's arsenal against ignorance and error; if over-relied upon, it could actually reduce the seeker closer to hapless than scholarly.

- AI's Impact on Problem-Solving Skills

Wu Jifan asserts that AI software like ChatGPT has transformed the way we approach problem-solving, utilize resources, and complete tasks. It has essentially made it so that tasks which were previously of a complex technical nature (ie programming) are now more accessible, narrowing the chasm in how technically competent individuals are. Anything and everything that used to take a lot of time, mature years or deep learning curve can be fetched instantly with the aid Artificial intelligence tools suggestions. As a result, tasks that once demonstrated professional expertise and problem-solving abilities may become trivial for professionals, reducing the level of accomplishment.

In programming, for example, previously it took a profound comprehension of code, debugging and algorithms to tackle a complex problem, ChatGPT can now give us automated solution reducing the process. This convenience speeds up development cycles and makes programming more accessible, but might also decrease the perceived value of problem solving of programmers. Relying on AI generated solution could end developing the critical thinking and creativity, as users find it comfortable to rely on AI instead of thinking a problem through for themselves.

This change both creates opportunities and challenges. This means that AI tools like ChatGPT on one hand are a productivity booster and time saver, one can channel the same saved time into strategic thinking, innovation thus leading to creativity. But on the flip side, there are also concerns that a world run by AI will erode technical skill and take away the personal fulfilment from solving hard problems. Thus, going forward it is important to find an equilibrium between taking advantages of the AI and not neglecting the human cognitive abilities that drive creativity and critical thinking.

- AI as a Gateway to Creativity

Lu Di states that artificial intelligence (AI), particularly models like GPT, represents a major advancement in technology and science, and thus marks significant progress. It is used across many industries and affects almost every part of our lives in the modern world. Colleges and universities often serve as institutions of knowledge dissemination and skill development, encouraging the use of AI so long as it is accomplished in an ethical way. Such institutions need to help learn AI as exploration, where the enabler for every young mind to think in depth and transform paper theories into practical solutions which build an imagination and foster creativity.

Educational institutions can adopt AI to enable students to see how these technologies work and prepare them for an increasingly AI driven future with widespread application in different professions. The difficulty to simply make sure that people will not misuse such technology, using AI as the source of uncredible information or a source which helps for bypassing processes. Instead of fighting the rise of AI, universities should prepare their students in using it ethically. He shared some interesting insight about the proliferation of AI in education, how we need to embed AI literacy in the curriculum material to help children understand what the strengths and limitations are of all technologies (not just AI) which will allow them to start leveraging technology as a complement to learning, rather than replacing critical thinking.

Universities need to drive academic integrity too, by ensuring students use AI ethically and in a way that prevents counterfeiting or superficial outputs. Teach students honesty and how to think for themselves, they will know what to do in a world where AI begins to express its innovations whilst being held accountable with ethics. In the end, it is up to universities to find a middle ground between giving in to AI's broader uses and instructing their students to complete tasks at that high level of academia when you know the technologies now being used only serve to augment human intellect instead if placing a cap over top of it.

In reflecting on AI policies within Hong Kong universities, team members offer varied viewpoints on integrating tools like ChatGPT into academic life. Li Jiefeng points out that educational institutions should adopt AI to boost creativity and access to information, but they must also be mindful of potential academic integrity issues like plagiarism. He suggests creating effective AI literacy programs to teach students and faculty about responsible use. Sharing a similar perspective, Yuan Kemeng emphasizes the need to balance AI's efficiency with nurturing critical thinking and problem-solving skills. He believes that while AI can enhance academic performance, students should also develop their own abilities and avoid becoming overly dependent on technology, which could hinder their progress in critical thinking (Facione, 2015). Wang Qing raises concerns about data integrity risks, noting that the vast amount of AI-generated information can lead to inaccuracies in research if not critically assessed (Binns et al., 2018). Wu Jifan discusses how AI has changed problem-solving methods, making complex tasks easier but possibly reducing the value placed on traditional skills (Brynjolfsson & McAfee, 2014). Lastly, Lu Di highlights the university's role in promoting ethical AI use and incorporating AI literacy into the curriculum to prepare students for a future where AI is prevalent (Zawacki-Richter et al., 2019). These insights collectively highlight the importance of developing policies that encourage responsible AI integration while protecting academic integrity and fostering critical thinking among students.